\def\bq{\begin{eqnarray}}
\def\eq{\end{eqnarray}}
\def\po{\begin{equation}}
\def\lo{\end{equation}}
\def\bm{\begin{pmatrix}}
\def\em{\end{pmatrix}}
\newcommand{\RR}{\mathbb R}

\def\abs{\hbox to 0.3cm{\hfil}}

\def\abs{\hbox to 0.3cm{\hfil}}

\def\n{\nonumber}
\def\ab{_{\alpha\beta}}
\def\ok{_{0k}}
\def\oi{_{0i}}

\def\n{\nonumber}
\def\ab{_{\alpha\beta}}
\def\ok{_{0k}}
\def\oi{_{0i}}

\def\oa{_{0\alpha}}
\def\ob{_{0\beta}}

\def\caik{\Gamma_{\;ik}^a}
\def\n{\nonumber}

\def\oa{_{0\alpha}}
\def\ob{_{0\beta}}
\def\ik{_{ik}}
\def\0i{_{0i}}
\def\0k{_{0k}}

\documentclass[12pt,a4paper]{amsart}
\usepackage{amsmath,amsthm,amstext,amssymb,a4}

\usepackage{epsfig}
\theoremstyle{plain}

\begin{document}
\author[]{M. Plaue \and M. Scherfner}
\address{M. Plaue and M. Scherfner: Institute of Mathematics, Technische Universit{\"a}t Berlin,
Str.~d.~17.~Juni 136, 10623 Berlin, Germany}
\author[]{L. A. M. de Sousa jr.}
\address{L. A. M. de Sousa jr.: Department of Mathematics, Universidade Federal do Estado do Rio de Janeiro, Av. Pasteur, 458, Urca, Rio de Janeiro, Brazil}

\title[]{On Spacetimes with given kinematical Invariants: Construction and Examples} 




\begin{abstract}
\noindent  We present a useful method for the construction of cosmological models by solving the differential equations arising from calculating the kinematical invariants (shear, rotation, expansion and acceleration) of an observer field in proper time description. As an application of our method we present two generalizations of the G\"odel spacetime that follow naturally from our approach.
\end{abstract}

\maketitle


\section{Introduction}
The construction of viable cosmological and astrophysical models often requires particular restrictions on the kinematic properties of those models. For example, parallax-free models must necessarily be shear-free \cite{hasse}.

Here the cosmological observer field is analyzed in proper time description to obtain expressions for the metric that depend explicitly on the kinematical invariants. This approach leads to a useful toolkit for constructing cosmological models with given kinematical properties. In addition, the analysis given here leads to a deeper understanding of the kinematical invariants and the relations between them. Of course, restrictions regarding the kinematical quantities also give rise to a limited range of possible matter models -- it is often not possible to give energy-momentum tensors representing simple matter like a perfect fluid with a particular combination of rotation, shear, expansion and acceleration. Recall for example the well-known shear-free fluid conjecture which has been proven for a number of special cases (cf. \cite{sopuerta}).

As an application of our method, we will construct a spacetime that generalizes the well-known G\"odel metric \cite{kurt} which represents a dust model with negative cosmological constant that has non-vanishing rotation, but vanishing shear, acceleration and expansion. It can be shown that this model contains closed timelike curves. Following \cite{hael}, p.\ 170 those properties suggest that the G\"odel spacetime is not very physical. In 1952, G{\"o}del \cite{kurt2} made some remarks about the necessity of constructing a solution with non-vanishing expansion, but without giving an explicit metric. In this paper, we will construct one generalization of the G\"odel spacetime for which the acceleration vanishes as well as another model that is parallax-free.

The authors wish to thank Professor Yuri Obukhov for many valuable suggestions while preparing the paper.

\section{Preliminaries}
We consider an $(N+1)$-dimensional smooth Lorentzian manifold $(M,g)$ of signature $(+,-,\dots,-)$. Given coordinates $(x^0,...,x^N)$, by $\partial_l$ we denote the partial derivative with respect to the $l^{\rm th}$ coordinate and $\nabla _j$ stands for the covariant derivative in direction of the $j^{\rm th}$ coordinate vector. Latin indices will take the values $0,\dots,N$, whereas greek indices will range from $1$ to $N$.
In the following we fix a timelike unit vector field $X^i$, the cosmological observer field. The usual decomposition of the covariant derivative of $X_i$ into irreducible parts (see for example \cite{colloq}) reads
\begin{eqnarray}
\nabla _kX_{i}=\omega_{ik}+\sigma_{ik}+\frac{1}{N}\Theta  P_{ik}
-\dot{X}_iX_k \label{feld}
\end{eqnarray}
with the antisymmetric part $\omega\ik$ (rotation), the symmetric traceless part $\sigma\ik$ (shear) and the trace $\Theta$ (expansion) itself. In detail, the parts read 
\begin{eqnarray}
\omega_{ik} &=& \nabla_{[i} X_{k]}-\dot{X}_{[i}X_{k]},\label{rot}\\
\sigma_{ik} &=& \nabla_{(i} X_{k)}-\dot{X}_{(i}X_{k)}-\frac{1}{N}\Theta  P_{ik},
\label{scher}
\end{eqnarray}
and
\begin{eqnarray}
\Theta  &=& \nabla _aX^a. \label{expansion}
\end{eqnarray}
Here the brackets (parentheses) denote the antisymmetric (symmetric) parts and
\begin{eqnarray}
P\ik=g\ik-X_iX_k \label{projektion}
\end{eqnarray}
is the projection tensor on the $N$-dimensional subspace perpendicular to $X^i$. The acceleration is given by
\begin{equation}
\dot X_i=X^a\nabla_aX_i.
\end{equation}

In addition we have
\begin{equation}
\dot X_aX^a=0,\quad\sigma_{ai} X^a=\omega_{ai}X^a=P_{ai} X^a=0. \label{zusatzgl1}
\end{equation}

\section{Proper time description of the observer field}
We adopt comoving coordinates with respect to the observer field such that $X^i=\delta^i_0$; this is also known as the proper time description. From $X^i$ being a timelike unit vector field we infer $1=X_aX^a=g_{ab} X^aX^b=g_{00}$. The first coordinate denotes the proper time of the cosmological observer; a fact which we will from now on emphasize by denoting $x^0$ with $t$.

In proper time description the components of the metric can be expressed by components of the observer field -- this is crucial to our analysis and we have:
\begin{equation}
X_i=g_{ai} X^a=g\oi
\end{equation}
and
\begin{equation}
P\ik =g\ik-g\oi g\ok \label{operator}.
\end{equation}

For the covariant derivative one has
\begin{eqnarray}
\nabla _kX_i &=& \partial_k X_{i}-\caik X_a\\
      &=& \partial_k X_{i}-\caik g_{0a}\n\\
      &=& \partial_k g_{0i}-\frac{1}{2}g^{ab}g_{0a} (\partial_kg_{ib} + \partial_i g_{kb} -\partial_b g_{ik} )\n\\
      &=& \partial_k g_{0i}-\frac{1}{2}\delta ^b_0 (\partial_kg_{ib} + \partial_i g_{kb} -\partial_b g_{ik} )\n\\
      &=& \frac{1}{2}(\partial_k g_{0i} - \partial_i g_{0k} + \partial_0 g_{ik}).\n\label{kou}
\end{eqnarray}

For the acceleration, one has in particular:
\begin{eqnarray}
\dot X_i &=& X^a\nabla _aX_i\\
	 &=&\nabla _0X_i\n\\
         &=& \partial_0 g_{0i} -{\frac{1}{2}}\partial_ig_{00}\n\\
         &=& \partial_0 g_{0i}\label{beschleunigung}.\n
\end{eqnarray}

In the same manner, we obtain similar expressions for the other kinematical quantities in proper time description:
\begin{eqnarray}
\Theta &=& \frac{1}{2}g^{ia}\partial_0 g_{ia}\\
&=&\frac{1}{2}\partial_0\left(\log{\rm det}
\left(g\ik\right)\right)\label{theta},\n\\
\omega _{ik} &=& \frac{1}{2}(g_{0k}\partial_0 g_{0i}-g_{0i}\partial_0 g_{0k}-\partial_k g_{0i}+\partial_i g_{0k}),
\label{rotation}\\
 \sigma _{ik} &=& \frac{1}{2}(\partial_0 g_{ik}-g_{0k}\partial_0 g_{0i}-g_{0i}\partial_0 g_{0k})
-\frac{1}{N}\Theta P\ik\n\\
&=& \frac{1}{2}\partial_0 P\ik-\frac{1}{N}\Theta P\ik\label{scherung}.
\end{eqnarray}

It can also be easily seen from (\ref{zusatzgl1}) that the kinematical quantities are characterized by their spatial components:
\po\dot X_0=0,\quad\omega _{0i}=\sigma _{0i}=P _{0i}=0\label{zusatzgl}.\lo

\section{Models with given kinematical invariants}
Equation (\ref{scherung}) can be solved for the projection tensor $P\ik=g\ik-g\oi g\ok$, which leads to
\begin{equation}
P\ik(t,x^\gamma)=\left(\Sigma\ik(t,x^\gamma)+P\ik(0,x^\gamma)\right)S^2(t,x^\gamma).
\end{equation}
with
\begin{equation}
S(t,x^\gamma)={\rm exp}\left(\frac{1}{N} \int\limits_0^t\Theta(\tau,x^\gamma)\,d\tau\right)
\end{equation}
and
\begin{equation}
\Sigma\ik(t,x^\gamma)=2\int\limits_0^t\frac{\sigma\ik(\tau,x^\gamma)}{S^2(\tau,x^\gamma)}\,d\tau.
\end{equation}
Denoting derivation with respect to $t$ by a dot, we have \begin{equation}\Theta=N\frac{\dot S}{S}\end{equation}  with $S(0,x^\gamma)=1$; $S(t,x^\gamma)$ is the so-called scale parameter.
\\\\
We interpret this equation as an evolution equation for the spatial components of the metric for a given ``start metric'' $g\ab(0,x^\gamma)$ and a given observer field $g_{0\alpha}(t,x^\gamma)$, expansion $\Theta(t,x^\gamma)$ and shear $\sigma\ab(t,x^\gamma)$ that drive the evolution:
\begin{eqnarray}
g\ab(t,x^\gamma)&=&g\oa(t,x^\gamma)g\ob(t,x^\gamma)+\label{evol}\\
&&S^2(t,x^\gamma)\left(g\ab(0,x^\gamma)-g\oa(0,x^\gamma)g\ob(0,x^\gamma)\right)+\n\\
&&S^2(t)\Sigma\ab(t,x^\gamma).\n
\end{eqnarray}

{\sl Remark:} The initial value ``$t=0$'' is not a specific coordinate value like e.g. a singularity. In fact, the construction is not valid for singular start metrics. One should regard $g\ab(0,x^\gamma)$ as the spatial metric at some arbitrary time in the history of the observer.

\section{Some special cases}
Whereas expansion, shear and the observer field itself enter the evolution equation directly, the rotation and acceleration impose additional constraints on the observer field.

\subsection*{Models with vanishing acceleration}
In the case of a vanishing acceleration, we have $\partial_0 g\oa=0$, which implies $g\oa(t,x^\gamma)=g\oa(0,x^\gamma)$, and (\ref{evol}) reduces to
\begin{eqnarray}
g\ab(t,x^\gamma)&=&S^2(t,x^\gamma)g\ab(0,x^\gamma)+\\
&&\left(1-S^2(t,x^\gamma)\right)g\oa(0,x^\gamma)g\ob(0,x^\gamma)+\n\\
&&S^2(t)\Sigma\ab(t,x^\gamma).\n
\end{eqnarray}

\subsection*{Irrotational models}
The rotation is calculated from the $g\oa$ components of the metric alone. The equation for an observer field with vanishing rotation can be solved by setting $g\oa=-h(t)\partial_\alpha \phi(x^\gamma)$. If the model also has vanishing acceleration, $h(t)$ is constant and the observer field is a gradient.

\subsection*{Parallax-free models}
Spacetimes with an observer field parallel to a conformal vector field are important since such models are precisely the parallax-free models. As was shown in \cite{hasse}, this condition holds if and only if $X^i$ is shear-free and the exterior derivative of $\dot X_i- \frac{\Theta}{N}X_i$ vanishes. The last condition can be met by assuming
\po\dot X_i=\frac{\Theta}{N}X_i-\partial_i f\label{para}\lo
for some function $f$.

Integrating (\ref{para}), and substituting into (\ref{evol}) with $\sigma\ab=0$, we have
\begin{eqnarray} g\ab(t,x^\gamma)&=&S^2(t,x^\gamma)g\ab(0,x^\gamma)+\\&&S^2(t,x^\gamma)F_\alpha(t,x^\gamma)F_\beta(t,x^\gamma)-\n\\&&S^2(t,x^\gamma)\left(F_\alpha(t,x^\gamma)g\ob(0,x^\gamma)+F_\beta(t,x^\gamma)g\oa(0,x^\gamma)\right)\n
\end{eqnarray}
with the functions
\begin{equation}
F_\alpha(t,x^\gamma)=\int\limits_0^t\frac{\partial_\alpha f(\tau,x^\gamma)}{S(\tau,x^\gamma)}\,d\tau.
\end{equation}

\subsection*{Toolkit}
Figure \ref{toolkit} can be used as a toolkit to construct spacetimes with specific kinematic properties. We have only listed shear-free models since a non-vanishing shear can be easily included via the functions $\Sigma\ab$. Also, we have notationally suppressed the dependence on the spatial coordinates $x^\gamma$.
\begin{figure}
\begin{tabular}{|c|c|}
\hline
 & \\
 & Spatial metric $g\ab(t)=g\ab(t,x^\gamma)$ \\
 & \\ \hline\hline
 & \\
General& $S^2(t)g\ab(0)+g\oa(t)g\ob(t)-S^2(t)g\oa(0)g\ob(0)$\\
 & \\ \hline
 & \\
$\dot X_i=0$ & $S^2(t)g\ab(0)+\left(1-S^2(t)\right)g\oa(0)g\ob(0)$\\
 & \\ \hline
 & \\
$\Theta=0$ & $g\ab(0)+g\oa(t)g\oa(t)-g\oa(0)g\oa(0)$\\
 & \\ \hline
 & \\
$\Theta=0,\dot X_i=0$ & $g\ab(0)$\\
 & \\ \hline
 & \\
$\omega\ik=0$ & $S^2(t)g\ab(0)+\left(h^2(t)-h^2(0)S^2(t)\right)\partial_\alpha\phi\partial_\beta\phi$\\
 & \\ \hline
 & \\
Parallax-free& $S^2(t)g\ab(0)+S^2(t)\left(F_\alpha(t) F_\beta(t)-F_\alpha(t) g\ob(0)-F_\beta(t) g\oa(0)\right)$\\
& \\ \hline
\end{tabular}
\caption{Table of shear-free cosmological models}
\label{toolkit}
\end{figure}
\section{Generalizations of the G\"odel spacetime}
We like to generalize the G\"odel spacetime $(\RR^4,\tilde g)$ equipped with the usual observer field to a shear-free model with non-vanishing expansion. Written in canonical (cartesian) coordinates $(t,x,y,z)$, the G\"odel metric reads:
\begin{equation}
\left(\tilde g\ik\right)=\bm1&0&e^{\sqrt{2}\omega_0x}&0\\0&-1&0&0\\e^{\sqrt{2}\omega_0x}&0&\frac{1}{2}e^{2\sqrt{2}\omega_0x}&0\\0&0&0&-1\em.
\end{equation}
In this coordinates, the observer field is already of the desired form, $X^i=\delta^i_0$. We take as a start metric
\begin{equation}
\left(g\ab(0,x^\gamma)\right)=\bm-1&0&0\\0&\frac{1}{2}e^{2\sqrt{2}\omega_0x}&0\\0&0&-1\em,
\end{equation}
and as the observer field we choose
\begin{equation}
\left(g\oa(t,x^\gamma)\right)=\bm0&f(t)e^{\sqrt{2}\omega_0x}&0\em.
\end{equation}
The time-dependent function $f(t)$ accounts for a possible non-vanishing acceleration.

In order to evolve this metric in a shear-free manner we use equation (\ref{evol}) with $\sigma\ab=0$. To simplify things, we assume that the scale parameter $S$ depends only on the cosmological time parameter $t$. In this way we arrive at
\begin{equation}
\left(g\ab(t,x^\gamma)\right)=\bm-S^2(t)&0&0\\0&\frac{1}{2}\left[S^2(t)\left(1-2f^2(0)\right)+2f^2(t)\right]e^{2\sqrt{2}\omega_0x}&0\\0&0&-S^2(t)\em.\label{evolgodel}
\end{equation}
\\
We like to model the matter content of this spacetime by a fluid (not necessarily perfect). Thus, the energy-momentum tensor takes the form
\begin{equation}
T\ik=-pg\ik+(\epsilon+p) X_iX_k+2q_{(i}X_{k)}+\pi\ik,
\end{equation}
with energy density $\epsilon$, isotropic pressure $p$, heat flow $q_i$ and anisotropic pressure $\pi\ik$.

The heat flow and anisotropic pressure have to satisfy the conditions $q_aX^a=0$, $\pi_{ai} X^a=0$ and $\pi^a_{\;a}=0$. The first condition directly implies $q_0=0$, whereas the second condition fixes $q_\gamma$ by inspection of the Einstein equation $R\ik-\frac{1}{2}Rg\ik-\Lambda g\ik=\kappa T\ik$ after solving for $\epsilon$ and $p$:
\begin{eqnarray}
q_1&=&\frac{\sqrt{2}\omega_0}{\kappa\left(2f^2(0)-1\right)}\frac{f(t)}{S^3(t)}\left(3\dot f(t)S(t)-\dot S(t)f(t)\right),\\
q_2&=&\frac{2}{\kappa}\frac{f(t)e^{\sqrt{2}\omega_0x}}{S^2(t)}\left(S(t)\ddot S(t)-\dot S(t)\right),\\
q_3&=&0.
\end{eqnarray}

\subsection*{Models with vanishing acceleration}
First we deal with the case of vanishing acceleration and set $f(t)\equiv1$. The evolved metric then reads
\begin{equation}
\left(g\ik\right)=\bm 1&0&e^{\sqrt{2}\omega_0x}&0\\0&-S^2(t)&0&0\\e^{\sqrt{2}\omega_0x}&0&\frac{1}{2}e^{2\sqrt{2}\omega_0x}\left(2-S^2(t)\right)&0\\0&0&0&-S^2(t)\em\label{godelacc}
\end{equation}
and the condition $\pi^a_{\;a}=0$ enforces
\begin{equation}
2 \omega_0^2\left(1-S^2(t)\right)-\ddot S(t)S(t)+2(\dot S(t))^2=0.\label{dgl}
\end{equation}
This equation can be reduced to the separable first order equation
\begin{equation}
(\dot S(t))^2=\omega_0^2\left(\left(\alpha-1\right) S^4(t)+2S^2(t)-1\right)\label{dgl1}
\end{equation}
which implies
\begin{equation}
\ddot S(t)=2\omega_0^2S(t)\left(\left(\alpha-1\right) S^2(t)+1\right) \label{dgl2}.
\end{equation}
Substituting (\ref{dgl1}) and (\ref{dgl2}) into the differential equation (\ref{dgl}) one sees that it is identically satisfied for any value of the constant $\alpha$. Furthermore, the equation can be solved in terms of elliptic integrals (the solution is one of the Jacobi elliptic functions).

\begin{figure}[htpb]
\begin{center}
\epsfig{figure=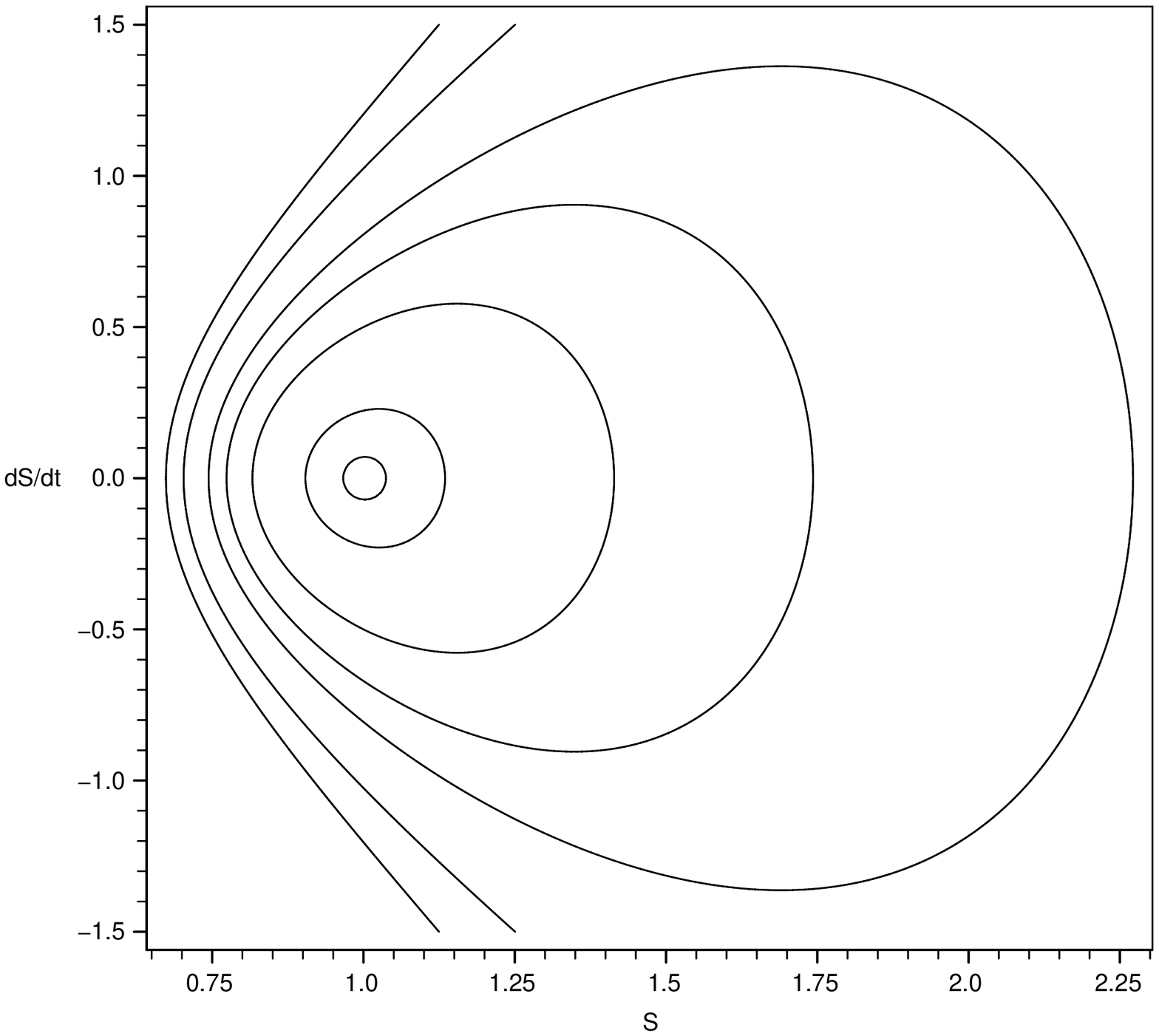,height=10cm,angle=0}
\caption[phase]{Phase portrait for the scale parameter $S(t)$ ($\omega_0=1$)}
\label{phase}
\end{center}
\end{figure}

However, evaluating (\ref{dgl1}) at $t=0$ one sees that $9\alpha\omega_0^2=9\dot S^2(0)=\Theta^2(0)$ and we must have $\alpha\geq0$ to obtain real valued solutions. For $\alpha=0=\Theta(0)$ one has $S(t)\equiv1$ and we recover the G\"odel metric. The phase portrait of the differential equation for different values of $\Theta(0)$ is shown in figure \ref{phase}: If $0<|\Theta(0)|<3|\omega_0|$, the solutions are periodic, whereas for $3|\omega_0|\leq|\Theta(0)|$ the spacetime expands/collapses indefinitely. The metric does not become singular at any point since $S(t)$ does not vanish for any initial value.

Furthermore, the model exhibits an interesting causal structure that we will discuss in the following section.

\subsection*{Parallax-free models}
We will now choose $f(t)$ such that the model is parallax-free. Calculating the exterior derivative of $\dot X_i-\frac{\Theta}{N}X_i$, we find that the observer field of our already shear-free model (\ref{evolgodel}) is parallel to a conformal vector field if and only if the following set of differential equations is satisfied:
\begin{eqnarray}
\ddot f(t)S(t)-\dot S(t)\dot f(t)&=&0,\\
\dot f(t)S(t)-\dot S(t) f(t)&=&0.
\end{eqnarray}
Together with the condition of a traceless anisotropic pressure we finally have $S(t)=e^{\frac{1}{3}\Theta t}=f(t)$ with the constant expansion $\Theta$.

The resulting metric takes the form
\begin{equation}
\left(g\ik\right)=\bm 1&0&e^{\sqrt{2}\omega_0x+\frac{1}{3}\Theta t}&0\\0&-e^{\frac{2}{3}\Theta t}&0&0\\e^{\sqrt{2}\omega_0x+\frac{1}{3}\Theta t}&0&\frac{1}{2}e^{2\sqrt{2}\omega_0x+\frac{2}{3}\Theta t}&0\\0&0&0&-e^{\frac{2}{3}\Theta t}\em,
\end{equation}
which is precisely the G\"odel metric for $\Theta=0$.

We compute some scalar invariants of this solution:
Rotation scalar and the norm of the acceleration are given by
\begin{eqnarray}
\omega_{ab}\omega^{ab}&=&2\omega_0^2e^{-\frac{2}{3}\Theta t},\\
\dot X_a\dot X^a&=&-\frac{2}{9}\Theta^2.
\end{eqnarray}

The heat flow and anisotropic pressure are non-zero:
\begin{equation}
-q_aq^a=\pi_{ab}\pi^{ab}=\frac{8}{9}\frac{\omega_0^2\Theta^2}{\kappa^2}e^{-\frac{2}{3}\Theta t}.
\end{equation}

Finally, the energy density and pressure are given by:
\begin{eqnarray}
\epsilon&=&\frac{1}{3\kappa}\left(3\omega_0^2e^{-\frac{2}{3}\Theta t}-\Theta^2-3\Lambda\right),\\
p&=&\frac{1}{3\kappa}\left(3\omega_0^2e^{-\frac{2}{3}\Theta t}+\Theta^2+3\Lambda\right).
\end{eqnarray}

A particular interesting case is $\Lambda=-\frac{\Theta^2}{3}$, where we have a stiff matter model $(\epsilon=p=\frac{\omega_0^2}{\kappa}e^{-\frac{2}{3}\Theta t})$ with non-vanishing anisotropic pressure and heat flow that is asymptotically a lambda vacuum for large cosmological times. In general, one must have $\Lambda\leq-\frac{\Theta^2}{3}$ to ensure that the cosmological observer measures an everywhere non-negative energy density $T_{ab} X^aX^b=\epsilon$.

\section{Closed causal curves}
Any function on a smooth closed curve $x^i(s)$ is periodic with respect to the curve parameter $s$. If the curve is contained in a chart of a Lorentzian manifold, this holds in particular for the coordinate function $x^0(s)$ and there exist maximal and minimal points of $x^0(s)$ for which $\frac{dx^0}{ds}=0$ holds.
Let $x^i(s)$ be such a closed curve that is also causal, i.e.
\po
g\ik\frac{dx^i}{ds}\frac{dx^k}{ds}\geq0
\lo
holds for all $s$. If $s_0$ is chosen such that $\frac{dx^0}{ds}=0$ holds 
(which is always possible by the arguments above) then
\po
\left.g\ik\frac{dx^i}{ds}\frac{dx^k}{ds}\right| _{s=s_0}=\left.g\ab\frac{dx^\alpha}{ds}
\frac{dx^\beta}{ds}\right| _{s=s_0}
\lo 
is negative if the matrix consisting of the $g\ab$ is negative definite -- which therefore is a sufficient condition for the absence of closed causal curves.

Following basic linear algebra a real square matrix $A=\left[a\ab\right]$ is negative
definite if $-A$ is positive definite. According to the Sylvester criterion, this holds if and only if the determinants of all leading principal minors $-A_\gamma$ (beginning with the element $-a_{11}$) are positive.

Examining the generalized G\"odel metric with vanishing acceleration (\ref{godelacc}), one immediately sees that this condition is met for $S^2(t)>2$. Thus, if the initial expansion compared to the initial rotation is large enough such that the solution of (\ref{dgl}) takes large enough values, the cosmological observer in the acceleration-free expanding generalized G\"odel spacetime ``escapes'' its causality violating epoch. If $\frac{27}{16}\left|\omega_0\right|<\left|\Theta(0)\right|<3|\omega_0|$, the observer periodically enters the chronology violating region of spacetime. If $3\left|\omega_0\right|\leq\left|\Theta(0)\right|$, this region is eventually avoided. Note that our argument does not show that the causality violating region is as large as $\left\lbrace S^2(t)\leq2\right\rbrace$. In particular, $\frac{27}{16}\left|\omega_0\right|<\left|\Theta(0)\right|$ may not be the best estimate for the initial values that give rise to causally well-behaved regions of this particular spacetime.

\section {Conclusion}
The construction of cosmological models with given kinematical invariants is an important task. On the one hand, it is an important problem in mathematical cosmology to examine possible connections between the kinematics of a cosmological model and its geometric properties like singularities and causality. On the other hand, astrophysical observations put constraints to the possible kinematical quantities of our universe. With the description of spacetime metrics presented here, these constraints can be directly accounted for.

The usefulness of our method is clearly demonstrated by the given examples. The parallax-free generalization of the G\"odel metric presented here seems to be more natural in terms of symmetry since we have just relaxed the property of a Killing observer field to an observer field parallel to a conformal vector field. The model with vanishing acceleration on the other hand has an interesting geometric structure with causality violating regions and periodic solutions.

\end{document}